\documentclass[a4paper,11pt]{article}
\usepackage{pos}

\title{Introducing Fermionic Link Models}
\author*[a]{Debasish Banerjee}
\author[b]{Emilie Huffman}
\author[c,d]{Lukas Rammelm\"{u}ller}

\affiliation[a]{Theory Division, Saha Institute of Nuclear Physics,\\
 1/AF Bidhannagar, Kolkata 700096, India}

\affiliation[b]{Perimeter Institute for Theoretical Physics,\\
 31 Caroline Street North, Waterloo, Ontario, Canada N2L 2Y5}
 
\affiliation[c]{Arnold Sommerfeld Center for Theoretical Physics (ASC), University of Munich,\\
 Theresienstr. 37, 80333 M\"unchen, Germany}
 
\affiliation[d]{Munich Center for Quantum Science and Technology (MCQST),\\
 Schellingstr. 4, 80799 M\"unchen, Germany}

\emailAdd{debasish.banerjee@saha.ac.in}
\emailAdd{ehuffman@perimeterinstitute.ca}
\emailAdd{lukas.rammelmueller@physik.uni-muenchen.de}

\abstract{Quantum link models (QLMs) are extensions of Wilson-type lattice gauge theories, and show 
rich physics beyond the phenomena of conventional Wilson gauge theories. Here we explore the physics
of $U(1)$ symmetric QLMs, both using a more conventional quantum spin-1/2 representation, as well as 
a fermionic representation. In 2D, we show that both bosonic and fermionic QLMs display the same physics. 
We then explore the models in 3D and find different behavior for the two QLMs. For the bosons, we see 
evidence for a quantum phase transition from a symmetry broken phase to a potential quantum spin liquid
phase. For the fermions, we identify not one but two distinct phases in addition to a symmetry broken
phase. We explore the symmetries of the ground state in the strong coupling limit, which breaks lattice
symmetries and examine the spectrum for both models.}

\FullConference{%
 The 38th International Symposium on Lattice Field Theory, LATTICE2021
  26th-30th July, 2021
  Zoom/Gather@Massachusetts Institute of Technology
}

\begin{document}
\maketitle

\section{Introduction}
  The Hamiltonian formulation of lattice gauge theories \cite{Kogut1974} (LGTs) has seen renewed interest in 
the context of recent theoretical and experimental developments. On the theoretical side, the introduction
and the exploration of quantum field theories in finite-dimensional Hilbert spaces \cite{Brower2003, 
Singh2019, Alexandru2019, Klco2018} have gathered a lot of momentum given the potential of realizing such 
models on quantum computers in the near future (see \cite{Preskill2018} for an overview). On the other 
hand, the last decades have seen astonishing developments in precise control of quantum systems, and 
engineering a variety of interactions between them \cite{Georgescu2014}. This has brought to life many 
toy models in tabletop quantum experiments \cite{Martinez2016, Klco2018a, Mil2020, Yang2020}. Moreover, 
such experiments have opened up the possibility of actually addressing physical problems, such as phases 
of frustrated magnets, quantum chromodynamics (QCD) at finite density, or real-time evolution of quantum 
many-body systems, where the sign problem presents an obstacle to simulations but not to experiments.

 One challenging aspect in this direction is the formulation of gauge theories on finite-dimensional Hilbert 
spaces. Compact gauge groups used to formulate LGTs non-perturbatively, as was first done by Wilson 
\cite{Wilson1974} to describe QCD, give rise to infinite-dimensional Hilbert spaces even locally. Naive 
truncation risks breaking gauge invariance. A theoretically appealing approach is to regulate this local 
infinity by gauge-invariant finite-dimensional operators, called quantum links \cite{Horn1981, Orland1989, 
Chandrasekharan1997}. For the simpler case of an $U(1)$ Abelian lattice gauge theory, these operators are 
quantum spins \cite{Wiese2021}, while for non-Abelian gauge theories, such as QCD, they can be chosen to 
be appropriate finite-dimensional operators \cite{Brower1997, Banerjee2017}. One particular representation  
using rishons (fermion bilinears) \cite{Brower1997} is particularly suited for realizing these models in 
analog quantum simulators \cite{Banerjee2012}. Furthermore, it is possible to realize the physics of Wilson
formulation in the quantum link approach by using large representations of the link operators 
\cite{Schlittgen2000,Zache2021}. One should note that different truncation schemes of the Hamiltonian formulation 
for gauge theories are also under active investigation \cite{Zohar2018,Raychowdhury2019,Davoudi2020}.

 Here we introduce a new class of models, generalizing the quantum link construction. After reviewing 
the formulation of the $U(1)$ quantum link model (QLM) with spin-$\frac{1}{2}$ quantum links, we show it is 
possible to realize a different gauge theory by choosing fermionic operators. For the square lattice, the 
theories realized with spin and with fermionic operators are identical to each other. However, for the cubic 
lattice the two models lead to differing physics, and give rise to rich phase diagrams. Various implications of this
construction are discussed.

\section{Plaquette operator in Abelian $U(1)$ LGTs}
 We first review the Hamiltonian formulation of Wilson LGTs. For an Abelian $U(1)$ gauge theory, the degree 
of freedom is a quantum rotor on the bond joining two lattice sites $x$ and $x+\hat{i}$, where the unit vector
$\hat{i}$ is a spatial direction. The rotor is characterized by an angle $\phi \in [0, 2 \pi[$, and the gauge 
field operators are $U_{x,\hat{i}} = \exp(i \phi_{x,\hat{i}})$ and 
$U_{x,\hat{i}}^\dagger = \exp(-i  \phi_{x,\hat{i}})$. 
The corresponding electric field is the angular momentum, $E_{x,\hat{i}} = -i\partial_{\phi_{x,\hat{i}}}$
and satisfies the following commutation relations:
\begin{equation}
 [E_{x,\hat{i}}, U_{y,\hat{j}}] = U_{x,\hat{i}} \delta_{xy} \delta_{ij},~~
 [E_{x,\hat{i}}, U_{y,\hat{j}}^\dagger] = -U_{x,\hat{i}}^\dagger \delta_{xy} \delta_{ij};~~
 [U_{x,\hat{i}}, U_{y,\hat{j}}^\dagger] = 0.
\label{eq:Comm}
\end{equation}
 It is convenient to work in the electric flux basis where $E_{x,\hat{i}}$ is integer-valued 
 $m_{x,\hat{i}} = 0, \pm 1, \pm 2, \cdots$. Thus the Hilbert space of even a single degree of freedom is infinite. 

\begin{figure}[!tbh]
 \includegraphics[scale=1.0]{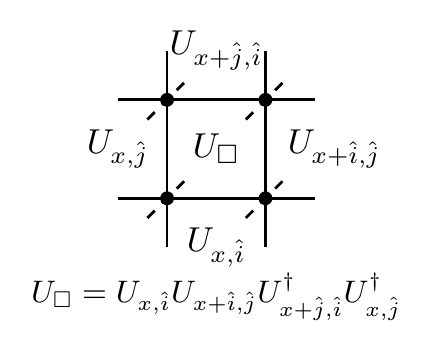}
 \includegraphics[scale=0.7]{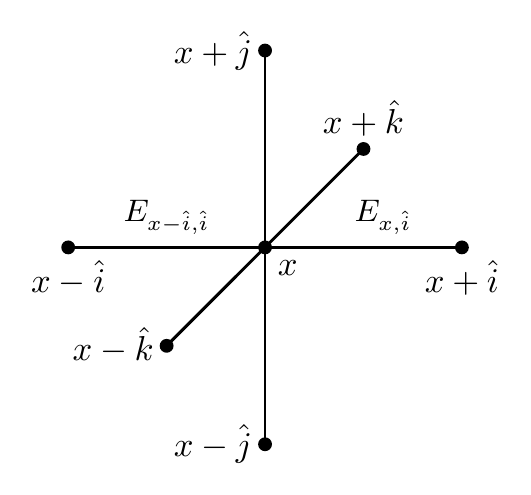}
 \includegraphics[scale=1.0]{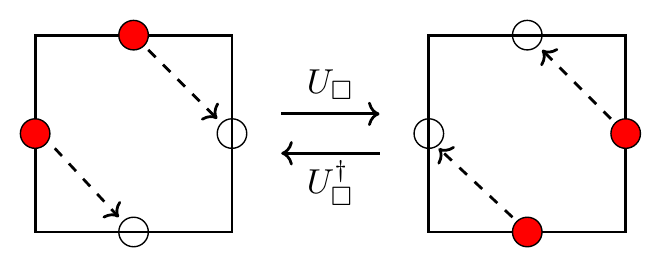}
 \caption{(Left) The plaquette operator on a cubic lattice. The dashed line represents the
third direction, perpendicular to the plane of the paper. (Middle) Six links touch a single site of the 3d
spatial lattice. Gauss' law thus involves 6 number operators, 3 in the forward and 3 in the backward links, 
each of which can be either empty or occupied. There are 20 states which satisfy the Gauss law.
(Right) The plaquette using fermionic operators can be understood a simultaneous hop of the
fermionic particles along the lines indicated, see text for details.}
 \label{fig:plaq}
\end{figure}

 For an extended lattice system, a configuration can be specified by the value of the electric flux at each 
 link. The gauge field operators $U_{x,\hat{i}} = L_{x,\hat{i}}^+$ and $~U_{x,\hat{i}}^\dagger = L_{x,\hat{i}}^-$ raise and lower the electric flux, respectively. The Hamiltonian for this system 
consists of the electric and the magnetic energy terms, the latter constructed from plaquette operators $U_\Box$
(with $e$ as the gauge coupling):
\begin{equation}
 H = \frac{e^2}{2} \sum_{x,i} E_{x,\hat{i}}^2 - \frac{1}{2 e^2} \sum_{\Box} (U_{\Box} + U^\dagger_{\Box});~~~~~~~~~~
 U_{\Box} = U_{x,i} U_{x+\hat{i},\hat{j}} U^\dagger_{x+\hat{j},\hat{i}} U^\dagger_{x,\hat{j}}.
\label{eq:Ham}
\end{equation}
The gauge invariance of the theory is due to the presence of local conserved charges
\begin{equation}
G_x = \sum_{i}(E_{x,\hat{i}} - E_{x-\hat{i},\hat{i}}), 
\label{eq:GLAW}
\end{equation}
the lattice Gauss law operators, which commute with the Hamiltonian $[G_x, H] = 0$. $G_x$ are the generators 
of gauge transformation $V = \prod_{x} \exp (i \theta_x G_x)$, with $\theta_x \in [0, 2\pi[$. The geometry
of the plaquette and the Gauss Law operator are shown in Fig. \ref{fig:plaq} (left) and (middle) respectively.  

  It is now interesting to consider how to realize the same local $U(1)$ symmetry using a finite-dimensional 
Hilbert space. One needs operators with the same commutations as Eq.(\ref{eq:Comm}), but nevertheless operating
on finite-dimensional Hilbert spaces. The simplest example is a quantum spin-$S$, having a a $(2 S + 1 )-$d 
Hilbert space, represented via $\vec{S} = (S^+, S^-, S^z)$. From elementary 
quantum mechanics we know that $[S^z , S^+] = S^+,~ [S^z , S^-] = -S^-$. Identifying the electric flux with the
$S^z$ component, the gauge fields are the raising and lowering operators of 
the electric flux, as before. The Hamiltonian of this QLM is given by Eq. (\ref{eq:Ham}), and the Gauss 
law by Eq. (\ref{eq:GLAW}), with the respective spin operators for the gauge and electric fields. 
Therefore, using quantum spin-$S$ as a degree of freedom on each link and identifying $U_{x,\hat{i}}=S^+_{x,\hat{i}},  
U^\dagger_{x,\hat{i}}=S^-_{x,\hat{i}}$, and $E_{x,\hat{i}} = S^z_{x,\hat{i}}$ realizes a $U(1)$ Abelian LGT
on $(2S+1)-$d local Hilbert space.

 It turns out that even the simplest non-trivial QLM has very rich physics. Realized at its most "quantum"
limit, the spin-$\frac{1}{2}$ representation of the $U(1)$ group has a local $2$-d Hilbert space on the links. The 
gauge symmetry is identical to that in the Wilson theory, since we only require the first two expressions
of Eq. (\ref{eq:Comm}) to prove the commutation $[H,G_{x}] = 0$. The most interesting consequence of the 
QLM construction is the non-unitarity of the gauge link operator: 
$[U_{x,\hat{i}}, U_{y,\hat{j}}^\dagger] = 2 E_{x,\hat{i}} \delta_{xy} \delta_{ij}$. In fact, this causes the 
QLMs to give rise to a wider class of physical phenomena than possible in Wilson-type gauge theories.
This theory in $(2+1)-$d gives rise to crystalline confined phases with fractionalized electric fluxes 
\cite{Banerjee2013,Banerjee2021}. It also has anomalous excited states, called quantum scars, which exhibit slow
thermalization \cite{Banerjee2020}.
 
  The rich physics of the QLM as compared to the Wilson theory motivates us to further our research into the class
of models which gives rise to newer phases of quantum gauge matter \cite{Flink2021}. 
A conceptually new model can be constructed using the $2$-d local Hilbert space of fermionic operators instead of 
spins. The presence or the absence of the fermion exhausts the two possibilities. Similar to the raising and lowering 
operators, the gauge fields are represented by creation or annihilation operators: 
$U_{x,\hat{i}} = c^\dagger_{x,\hat{i}},~U^\dagger_{x,\hat{i}} = c_{x,\hat{i}}$. Moreover, the generator of the
gauge transformations is the number operator, which is also identified with the electric flux: 
$E_{x,\hat{i}} = (n_{x,\hat{i}} - \frac{1}{2})$. Of course, this entire construction only works since these 
operators also satisfy the commutation relations amongst each other as the quantum spins:
\begin{align}
[n_{x,\hat{i}}, c^\dagger_{y,\hat{j}}] &= c^\dagger_{x,\hat{i}} \delta_{xy} \delta_{ij};~
[n_{x,\hat{i}}, c_{y,\hat{j}}] = c_{x,\hat{i}} \delta_{xy} \delta_{ij};~
[c^\dagger_{x,\hat{i}}, c_{y,\hat{j}}] = 2 E_{x,\hat{i}} = 2 (n_{x,\hat{i}} - \frac{1}{2}); \\
\{c_{x,\hat{i}}, c_{y,\hat{j}} \} &= \{c^\dagger_{x,\hat{i}}, c^\dagger_{y,\hat{j}} \} = 0;~~
\{ c^\dagger_{x,\hat{i}}, c_{y,\hat{j}}  \} = \delta_{xy} \delta_{ij}.
\end{align}
Because of the anti-commutation relations, naively one can expect that the physics of this gauge theory
will be very different from the one formed using quantum spins (i.e. bosonic links).

\section{The plaquette model with fermionic links}
  In terms of the fermionic operators, the plaquette and Gauss' law are:
\begin{equation}
 U_{\Box} = c^\dagger_{x,\hat{i}} c^\dagger_{x+\hat{i},\hat{j}} c_{x+\hat{j},\hat{i}} c_{x,\hat{j}};~~
 G_x = \sum_{i} \left( n_{x,\hat{i}} - n_{x-\hat{i},\hat{i}} \right).
 \label{eq:GLfermi}
\end{equation}
  Even though it might look onerous at first sight, the plaquette operator can be interpreted as a 
simultaneous hop of two fermions as shown in the right panel of Fig.~\ref{fig:plaq}. However, the hopping can only occur
if the shaded sites are occupied and their directed neighbors are empty. Thus, among the six hoppings one can
imagine, only two are actually allowed. 

 As long as we construct the Hamiltonian from gauge-invariant operators, the resulting model will retain
the gauge invariance. Moreover, the electric flux energy term contributes a constant,
and hence we omit it from the Hamiltonian. However, it is always possible to add a background electric field 
term which couples to the total electric flux. This can be considered as a self-adjoint extension, which we
disregard for the moment. Instead, we consider a Rokhsar-Kivelson-like term, also considered for the usual
QLM realized with spins:
\begin{equation}
 H_{\rm F} = -J \sum_{\Box} (U_\Box + U^\dagger_\Box) + \lambda \sum_{\Box} (U_\Box + U^\dagger_\Box)^2.
 \label{eq:FHam}
\end{equation}

\begin{figure}[!tbh]
 \begin{center}
 \includegraphics[scale=0.8]{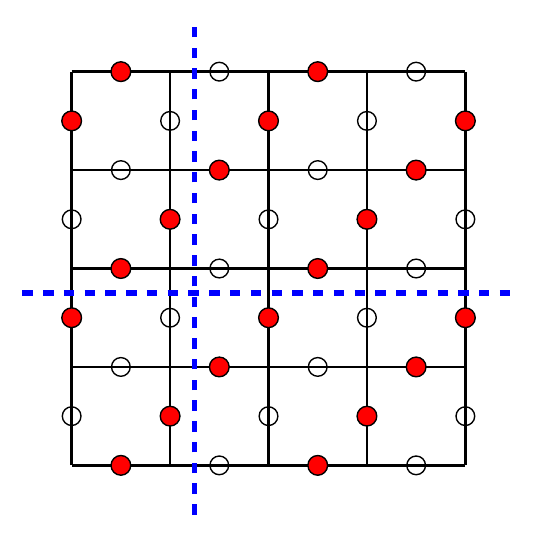}
 \includegraphics[scale=0.8]{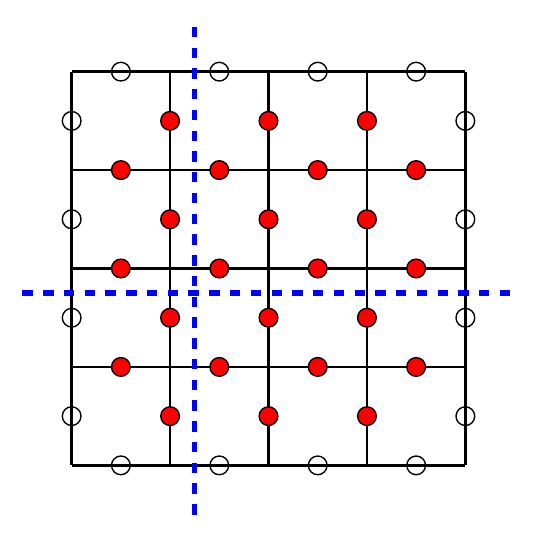}
 \end{center}
 \caption{ Examples of configurations with zero (left) and finite (right) winding number states. The winding number
 equals half of the difference between the occupied and empty links touched by the vertical (horizontal) line in 
 the respective directions. These are topological quantities, and the exact location of the line is irrelevant as 
 long as the line winds across the lattice.}
 \label{fig:Wind}
\end{figure}

 The Hamiltonian satisfies the usual lattice point group symmetries, such as lattice translation, rotation, parity,
as well as charge conjugation, a discrete $\mathbb{Z}(2)$ symmetry. A detailed discussion
of the lattice symmetries has implications on the understanding of its phase diagram and is discussed in depth
in \cite{Flink2021}. The only other global internal symmetry of the model is the $\mathbb{U}(1)^d$
generated by large gauge transformation (winding numbers). In 2 spatial dimensions, these operators are:
\begin{equation}
{W}_{x} = \frac{1}{{L}_{x}} \sum_x ({n}_{r=(x,y_0),\hat{j}} - \frac{1}{2});~~~~~~~~~~
{W}_{y} = \frac{1}{{L}_{y}} \sum_y ({n}_{r=(x_0,y),\hat{i}} - \frac{1}{2}).
\end{equation}
Every state in the electric flux basis can be characterized by a specific winding number, and two such examples are 
shown in Fig. \ref{fig:Wind}. The Hamiltonian commutes with ${\rm W}_{\rm x}, {\rm W}_{\rm y}$, and they can be used
to block-diagonalize the Hamiltonian. 

\section{From two to three spatial dimensions}
 In this section, we argue that the proposed model has identical physics with spin-$\frac{1}{2}$ model in two spatial
dimensions, but not in three. To understand the difference, we ask if it is possible for the fermions to interchange
their spatial positions. If this is allowed, then the fermion anti-commutation properties will lead to additional negative sign factors and potentially change the physics.
Otherwise, the physics will be the same, since the Hamiltonian, the Gauss law and the Hilbert space are identical
for both formulations.

\begin{figure}[!tbh]
 \begin{center}
 \includegraphics[scale=0.6]{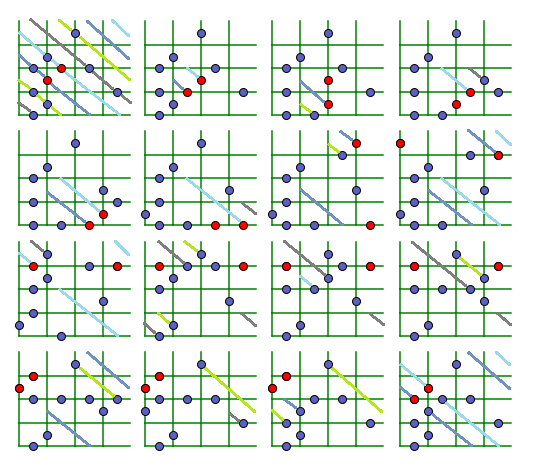}
 \end{center}
 \caption{ From left to right, top to bottom:  A set of plaquette hops to make the fermions (shaded in red) exchange 
 their positions with their neighbours (shaded in blue) on the same diagonal lines (blue and cyan).
 The nature of the plaquette interactions moves both the red fermions, and at the end
 of the sequence \emph{both} pairs of fermions exchange positions, even in finite volume. There are no unpaired
 negative signs due to Pauli exclusion principle, and the physics is the same as in the bosonic version.}
 \label{fig:worldline2d}
\end{figure}

 We will present a geometrical proof of the statement. Note from Fig.~\ref{fig:plaq} (right) that the plaquette 
interaction causes the fermions to move diagonally along the dotted lines, which pass through the mid-point 
of the neighbouring links. As indicated in Fig.~\ref{fig:worldline2d}, the motion of the fermions is restricted to 
1-d non-overlapping lines, which wrap across the boundary, but do not intersect. As a result, it is 
impossible for two fermions to interchange places, except by going across the boundary. However, another key feature
of the interaction is that it involves a \emph{simultaneous} hop of a fermion on an adjacent line. Therefore, 
for every instance when two fermions on a line interchange their positions, another exchange happens between
two other fermions in one of the two adjacent lines. Consequently, there is never a net negative sign encountered.
Thus the physics of the model with spins and fermions in two dimensions are identical.  

\begin{figure}[!tbh]
 \begin{center}
 \includegraphics[scale=0.4]{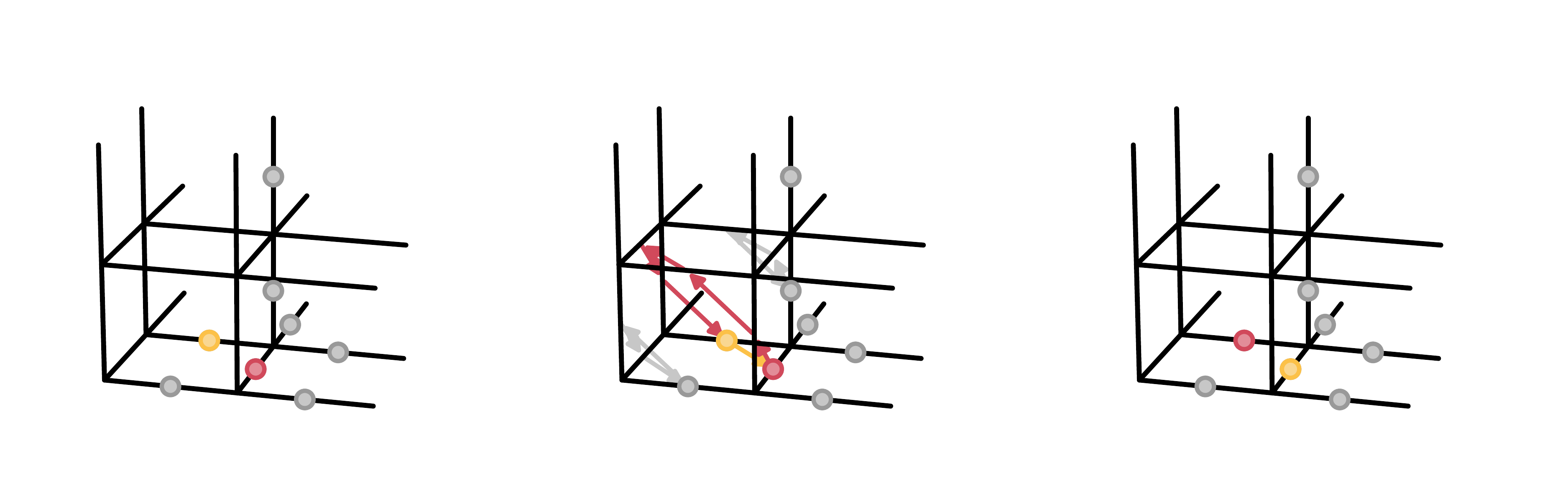}
 \end{center}
 \caption{ From left to right: Two fermions colored in red and yellow have been chosen among the inactive
 (grey) fermions, whose positions are exchanged through a series of flips. Finally, the red and yellow have
 switched places, keeping all the grey ones at their original positions. This is an example of a configuration
 which has an overall negative sign compared to the spin version.}
 \label{fig:worldline3d}
\end{figure}

 In three dimensions, this argument does not work. Drawing the paths on which the fermions can move, one
traces out a set of non-intersecting planes, as shown in Fig.~\ref{fig:worldline3d}. On the plane, it is easily
possible for a pair of fermions to interchange their positions (shown in the figure as red and yellow) keeping
the position of all other fermions unchanged (shown in grey). We have verified the existence of paths
which have an extra negative sign due to this interchange. Since such negative signs are absent in the corresponding
model formulated with quantum spin-$\frac{1}{2}$, it is reasonable to expect that the two models show different
physics.

\section{Physics of the 3d model and Outlook}
 Motivated by the prospect of investigating novel phases in this class of models, we have carried out large
scale exact diagonalization (ED) studies for both the bosonic and the fermionic model in $(3+1)-$d \cite{Flink2021}.
The complete Hamiltonian we consider is the one in Eq.~(\ref{eq:FHam}), with the Gauss law as in Eq.~(\ref{eq:GLfermi}).
ED calculations are exponentially expensive, but nevertheless by restricting to the Gauss law sector and by exploiting 
the global symmetries extensively, we have been able to diagonalize systems of up to 72 particles (spins) and gain some 
insight into the phase diagram using finite size scaling (FSS). 

 On very general grounds, we expect that the phase of both the models will be similar for large and 
negative $\lambda$. The $\lambda$-term counts whether the plaquettes are
flippable (if the position of the fermions in the plaquette are in the preferred two of the six total possibilities). 
For large negative $\lambda$ therefore, this term dominates over the fermion-hopping. In the
absence of fermionic hopping, the anti-commutations cannot play a role, and the two models would have similar
physics. 

\begin{figure}[!tbh]
 \begin{center}
 \includegraphics[scale=0.3]{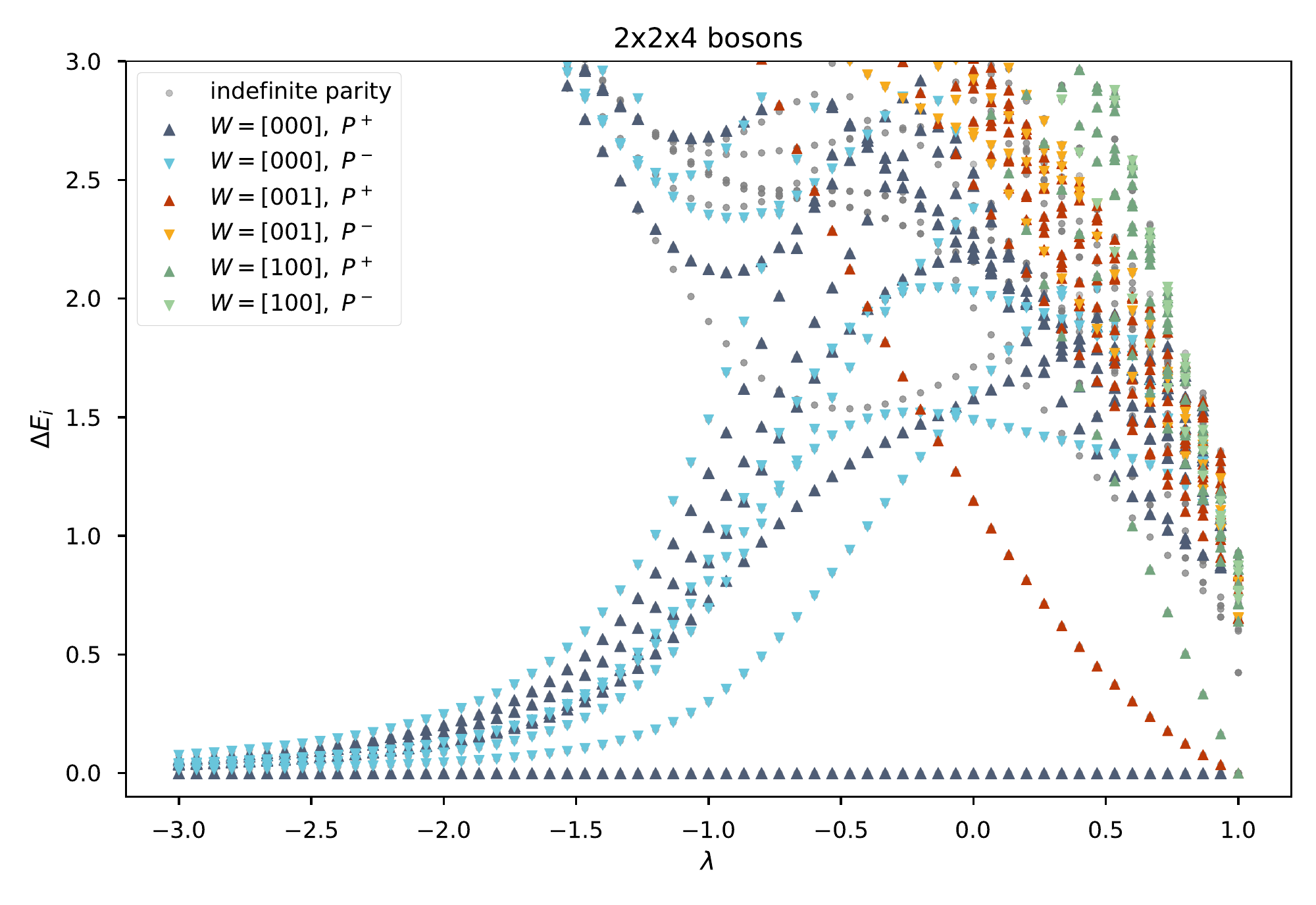}
 \includegraphics[scale=0.3]{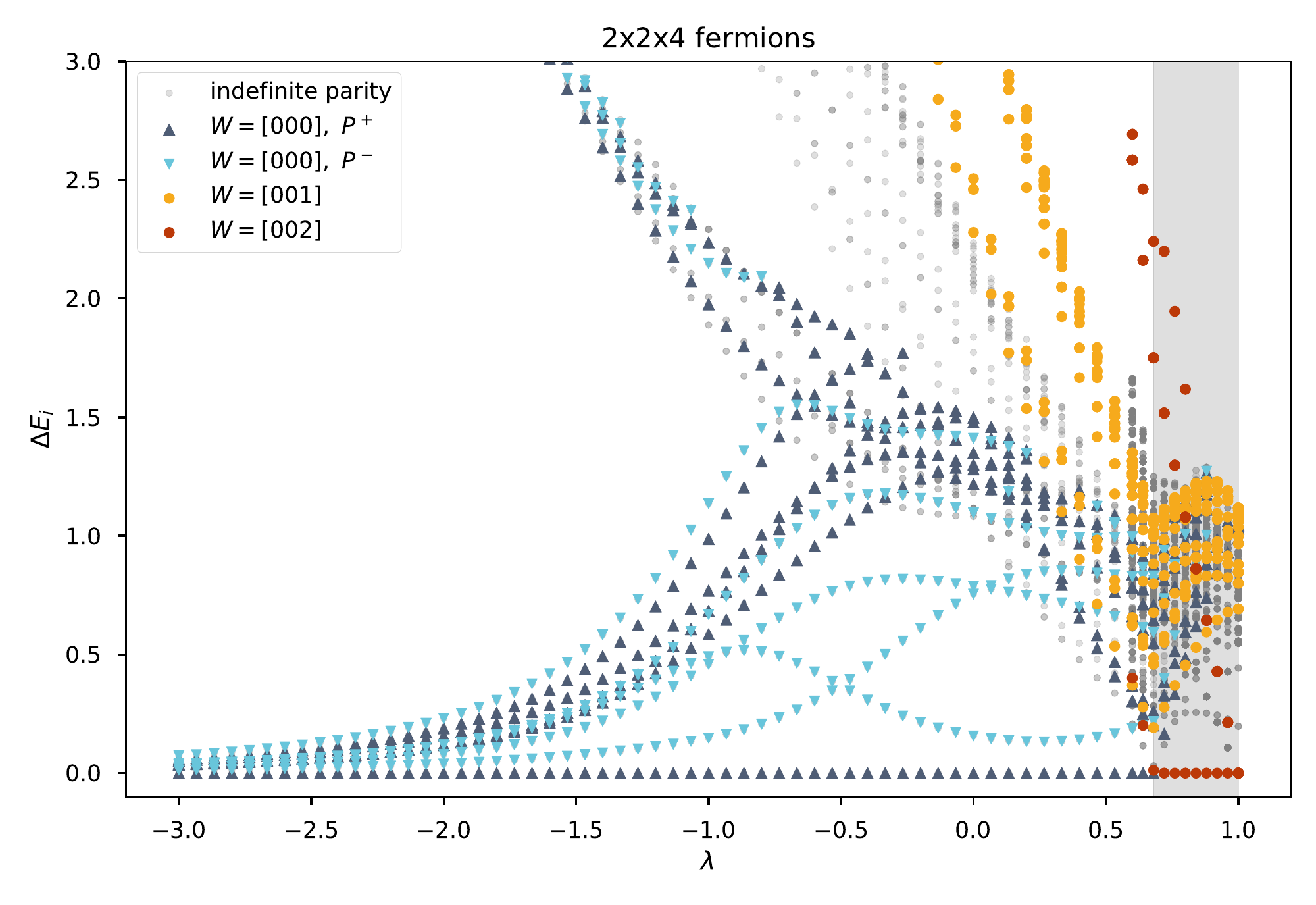}
 \end{center}
 \caption{Gap of the lowest energy eigenstates realized with spin (left) and fermion (right) operators.}
 \label{fig:phaseDiag}
\end{figure}

 In Fig.~\ref{fig:phaseDiag} we have plotted the lowest 15 eigenvalues of the Hamiltonian as a function of $\lambda$
 both the bosonic (left) and the fermionic (right) models. In region of large negative $\lambda$, the spectra
 are similar, even quantitatively. A closer analysis of the energy levels indicate a spontaneous symmetry breaking (SSB)
 of the rotational symmetry, discussed in detail in \cite{Flink2021}. In between $-1 < \lambda < 0$, both models 
 show different behaviour of the first excited state. For the fermionic model there is a level crossing, while 
 for the bosonic model the sign of the SSB disappears. The FSS is consistent with the
 existence of a Coulomb phase in this region \cite{Flink2021}. On increasing $\lambda \rightarrow 1$, both models
 flow to the Rokhsar-Kivelson point, which brings in the fluxes from the excited states. The fermionic model seems
 to have a narrow region for $\lambda <  1$ where the flux is condensed \cite{Flink2021}. 

 In addition to the physics issues explored in the \cite{Flink2021}, there are several very intriguing conceptual ideas about this
 particular model which branch out from this work. An almost obvious question is how to explore the model on larger 
 lattices in three spatial dimensions. Quantum Monte Carlo is not efficient for the bosonic model, and for the 
 fermionic model there is even a sign problem. On the experimental side, our reinterpretation of the plaquette interaction 
 could be an important way of realizing this on cold atom quantum simulators, an issue which has remained challenging. We may also ask if dynamical signatures about the thermalization of this model in two dimensions manifest in three dimensions. 
 
 Another completely different set of questions could result in further development of the ideas presented here.
 Since we are working with fermions, how does one reach the Kogut-Susskind-Wilson limit? If there is a phase with
 large correlation length, can we take the continuum limit using dimensional reduction? More importantly, what
 would the low-energy effective theory with fermionic photons look like? Would this idea generalize to non-Abelian 
 gauge theories? We believe that answering even some of these questions would increase our understanding of 
 quantum field theories at the frontiers.

\bibliographystyle{JHEP}
\bibliography{refFqlm}
%\begin{thebibliography}{99}
%\bibitem{...}
%\end{thebibliography}

\end{document}